\documentclass{article}
\usepackage[utf8]{inputenc}
\usepackage{amsmath}
\DeclareMathAlphabet{\mathpzc}{OT1}{pzc}{m}{it}
\usepackage{amsfonts}
\usepackage{bm}
\usepackage{graphicx}
\usepackage{float}
\usepackage[font={footnotesize}, labelfont={bf}]{caption}

\newcommand{\bi}[1]{\bm{\mathit{#1}}}
\newcommand{\Kbar}{\overline{K}}
\newcommand{\K}{\mathpzc{K}}

\newcommand{\area}{\mathcal{A}}
\newcommand{\curve}{\mathcal{C}}
\newcommand{\sat}{\mathcal{S}}
\newcommand{\zb}{z_b}
\newcommand{\zt}{z_t}
\newcommand{\ts}{\textsuperscript}
\newcommand{\p}{\partial}
\newcommand{\pt}{\frac{\partial}{\partial t}}
\newcommand{\px}{\frac{\partial}{\partial x}}
\newcommand{\py}{\frac{\partial}{\partial y}}
\newcommand{\dd}{\, \textrm{d}}

\begin{document}

\title{Equations for Long-Range Groundwater Flow and Water Table Evolution}
\author{Mark Baum}
\date{March 5, 2019}
\maketitle

\begin{abstract}
These are notes that I compiled while studying the equations of long-range groundwater flow for my first paper \cite{Baum20}. By ``long-range," I mean horizontal distances that are significantly greater than the vertical thickness of the aquifer, in addition to some other assumptions discussed below. None of this material constitutes original development of important new equations, but it might be useful to someone learning about simplified models of groundwater flow for numerical solution.

I start from Darcy's Law, reminding myself how to formulate the vertically averaged conservation law for groundwater flow. Then I discuss some ideas about how to solve the flow equation numerically. After that, I go through some cases where the hydraulic conductivity is a simple function of the vertical coordinate and the the steady-state flow equation can be simplified. Finally, there is a very short discussion of the transient case.
\end{abstract}

\tableofcontents
\newpage

\section{Governing Equation, From Darcy's Law}

\subsection{Dupuit Approximation and Vertical Integration}

The Dupuit approximation (see \cite{Bear72}, Chapter 8) defines volumetric groundwater flux $\bi{q}$ in an unconfined aquifer when the flow is essentially horizontal. In this case, the piezometric head $\psi$ is vertically uniform and the usual form of Darcy's law, $q = -K \nabla \psi$, can be expressed in terms of the water table $h$, which is the height of the free surface at the top of the saturated zone.
\begin{equation}
	\bi{q}(x,y,z) = -K(x,y,z) \nabla h(x,y) \label{q} \, ,
\end{equation}
where $\bi{q}$ is water flux (m/s) and $K = \rho g k/\mu$ is the isotropic hydraulic conductivity (m/s). We can assume $\nabla h$ is vertically uniform because $\psi=h$ along any vertical line for a hydrostatic aquifer where there is no vertical flow. If $(K_{xy}/K_z)\nabla h \ll 1$, where $K_{xy}$ is a representative horizontal conductivity and $K_z$ is the vertical conductivity, the approximation is reasonably accurate \cite{Bear72}.

Equation \ref{q} can be vertically averaged over the thickness of the aquifer where flow is occurring, the saturated zone, which is a vertical thickness denoted $\sat$. Because $\nabla h$ is vertically uniform, it comes out of the integral.
\begin{align}
	\int_\sat \bi{q}(x,y,z) dz = \bi{Q}(x,y) &= \int_\sat -K(x,y,z) \nabla h(x,y) \dd z \\[1em]
	&= - \nabla h(x,y) \int_\sat K(x,y,z) \dd z \\[1em]
	&= - T \Kbar \nabla h(x,y) , \label{Q}
\end{align}
where $\Kbar$ is the vertically averaged (but still spatially varying) conductivity of the aquifer, $T$ is the depth of the vertical average,
\begin{equation}
	\Kbar(x,y) = \frac{1}{T}\int_\sat K(x,y,z) \dd z \, ,
\end{equation}
and $\bi{Q}$ is the vertically integrated volume flux. If we know that the bottom of the aquifer occurs at $h=0$, then $T=h$ and
\begin{equation}
	\Kbar(x,y) = \frac{1}{h(x,y)}\int_\sat K(x,y,z) \dd z \, .
\end{equation}

\subsection{Conservation Law}

After integrating the usual Darcy flux over the vertical dimension, Equation \ref{Q} defines the volume of water flowing through a unit horizontal width of aquifer per unit time. It is the \emph{vertical} integral of \emph{horizontal} fluxes and a function of $x$ and $y$ alone. The Dupuit assumption removes the vertical coordinate from the flow.

In an arbitrary column of an aquifer with horizontal area $\area$, the total volume of water varies according to the volume flux at the surface defining its boundary and with internal sources $F$ (units of $[L/T]$). If the column is a cylinder, for example, $\area$ is the area of the circular caps and its boundary surface is the curved body of the cylinder. With the vertical integration already wrapped into $\bi{Q}$, the volume flux into the column is an integral of $\bi{Q}$ over the two-dimensional curve $\curve$ defining the column's horizontal boundary, the edge of its footprint.
\begin{equation}
	\pt\iint_\area  V(x,y)  + \oint_\curve \bi{Q}(x,y) \cdot \hat{n} \, dl = \iint_\area F(x,y) \, ,
\end{equation}
where $V$ is water volume per unit area, $\hat{n}$ is the unit vector normal to $\curve$ (pointing outward), and $dl$ is a tiny integration length along $\curve$. Using the divergence theorem to convert the line integral of $Q$ to a surface integral of its divergence, 
\begin{align}
	\pt\iint_\area  V(x,y)  + \iint_\area \nabla \cdot \bi{Q}(x,y) = \iint_\area F(x,y) \\[1em]
	\iint_\area \left[ \pt  V(x,y)  + \nabla \cdot \bi{Q}(x,y)  - F(x,y)  \right] = 0  \, . \label{div}
\end{align}
Because $\area$ is an arbitrary area and Equation \ref{div} must be satisfied for any choice of $\area$, the integrand must be zero everywhere, leading to the differential form of the volume conservation law,
\begin{equation}
	\pt  V = - \nabla \cdot \bi{Q} + F \, , \label{cons1}
\end{equation}
where $V$ is water volume per unit aquifer area and $F$ is volume source term per unit aquifer area, and everything is still a function of $x$ and $y$.

The conservation law above is conceptually clear but abstract and not particularly useful. We'd like to describe the evolution of the height of the saturated zone, or the water table $h$. The flux $Q$ depends on the gradient of $h$, so the first term on the right above will involve something like the Laplacian of $h$, $\nabla^2 h$. To solve the equation for $h$ in different scenarios, we need to express the time derivative on the left side in terms of $h$ as well. This can be done by relating the volume $V$ to the porosity, then applying a calculus trick. 

In the saturated zone ($z \leq h$) we assume water occupies all of the pore space. Above the water table, we assume there is no water. So, if porosity $n(z)$ is a function of depth, the volume of water per unit area of a column of aquifer is the sum of all pore space below the water table,
\begin{equation}
	V = \int_\sat n(z) \dd z \, , \label{V}
\end{equation}
with units of length (volume per area). This is the water column for a specific horizontal position in the aquifer. The total volume of water in the aquifer would be the integral of $V$ over $x$ and $y$. Equation \ref{V} is still not particularly helpful for solving the conservation law, but it can be rearranged with the chain rule. If the integral is carried out from the bottom of the aquifer $\zb$ to the top of the saturated zone, which is the water table $h$,
\begin{align}
	\pt V &= \pt \int_{\zb}^h n(z) \dd z \\
	&= \pt \left[ N(z) \right] \bigg|_{\zb}^{h} \\[1ex]
	&= \pt \left[ N(h) - N(\zb) \right] \\[1ex]	
	&= \frac{\p N(h)}{\p z} \frac{\p z}{\p h} \frac{\p h}{\p t} - \frac{\p N(\zb)}{\p z} \frac{\p z}{\p \zb}  \frac{\p \zb}{\p t} \\[1ex]
	&= \frac{\p N(h)}{\p z} \cdot 1 \cdot \frac{\p h}{\p t} - \frac{\p N(\zb)}{\p z}  \cdot 1 \cdot 0 \\[1ex]
	&= n(h) \frac{\p h}{\p t} \, .
\end{align}
Plugging this result into the general conservation equation (Eq. \ref{cons1}) along with the expression for $\bi{Q}$ from Equation \ref{Q}, we have
\begin{equation}
	n(h) \frac{\p h}{\p t} = \nabla \cdot \left( \Kbar T \nabla h \right) + F \, ,
	\label{cons2}
\end{equation}
or, if the bottom of the aquifer (below which flow is zero) is always at $h=0$, so that $T=h$,
\begin{equation}
	n(h) \frac{\p h}{\p t} = \nabla \cdot \left( \Kbar h \nabla h \right) + F \, .
\end{equation}
Most generally though, we can leave the depth integration of $K$ in the conservation equation explicitly,
\begin{equation}
	n(h) \frac{\p h}{\p t} = \nabla \cdot \left( \nabla h \int_{\zb}^h K \dd z \right) + F \, ,
	\label{cons3}
\end{equation}

Equation \ref{cons2} is a useful form of the conservation law in the sense that it only involves $h$ and other known or prescribed parameters, $K$, $F$, and $T$. We can't work directly with $V$ and $Q$. Equation \ref{cons2} is not easy to solve in the general case though, even numerically. $K$ may be spatially variable and $\Kbar$ might be temporally variable through its dependence on the coordinates of the saturated zone. Additionally, if $\int K$ is not easy to find analytically, then $\Kbar$ must be computed numerically whenever $h$ changes.

\subsection{For Numerical Solution}

It might be easier to solve Equation \ref{cons3} numerically if it's first written in another form. If $\Kbar$ is not analytically available, the depth-integrated conductivity would have to be evaluated numerically. Writing Equation \ref{cons3} in one horizontal dimension for ease,
\begin{equation}
	n(h) \frac{\p h}{\p t} = \px \left( \frac{\p h}{\p x} \int_{\zb}^h K \dd z \right) + F \, .
\end{equation}
Using the product rule,
\begin{equation}
	n(h) \frac{\p h}{\p t} = \frac{\p h}{\p x}\px \int_{\zb}^h K \dd z + \frac{\p^2h}{\p x^2} \int_{\zb}^h K \dd z + F \, . \label{consprod}
\end{equation}
The derivatives of $h$ can be evaluated with finite differences, spectral methods, or whatever else. The integral of $K$ can be done numerically for each of the spatial nodes with a quadrature algorithm or, in the best case, analytically. That leaves only the derivative of the integral in the first term on the right,
\begin{equation}
	\px \int_{\zb}^h K \dd z \, . \label{dKint}
\end{equation}

\begin{figure}[ht]
	\centering
	\includegraphics[scale=0.75]{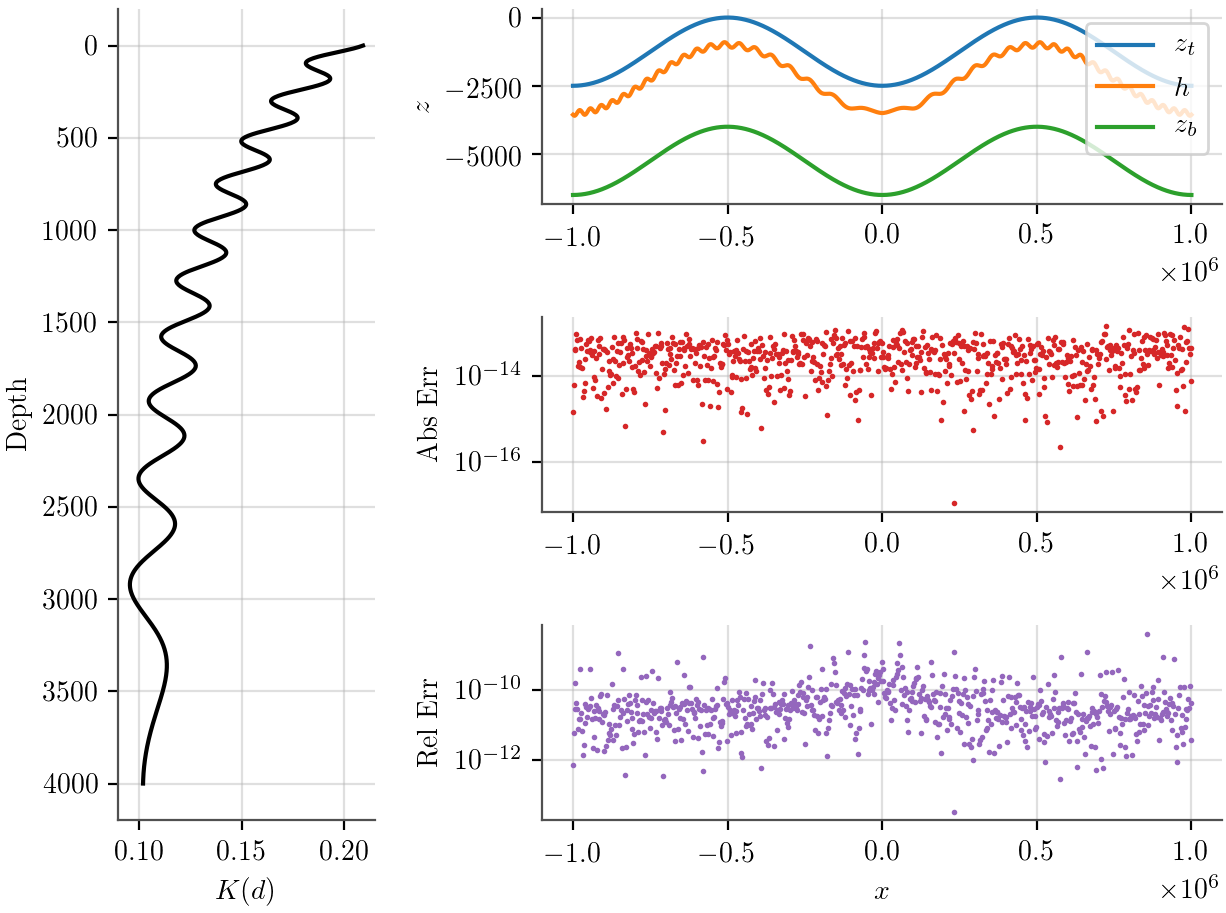}
	\caption{A test of Equation \ref{dKintind}. The tall panel on the left shows the hydraulic conductivity profile with depth. The top panel on the right shows the top of the aquifer, the water table, and the bottom of the aquifer. The middle panel shows the absolute error between a direct evaluation of Expression \ref{dKint} and the indirect evaluation represented in Equation \ref{dKintind}. The bottom panel shows the relative error. The derivatives are evaluated with a five-point centered difference and the integrals with \texttt{scipy.integrate.quad}. The test shows that, even with a wiggly conductivity profile and a wavy groundwater table, Equation \ref{dKintind} is equivalent to Equation \ref{dKint} to numerical accuracy.}
	\label{fig:dKint}
\end{figure}

Evaluating this term directly could be quite computationally expensive. Fortunately, if we make two assumptions, the derivative of this integral can be greatly simplified. First, we assume $K$ depends only on depth. Second, we assume the thickness of the aquifer is uniform even if the surface height is not, so that $\zt - \zb$ is constant. Writing the integral in terms of depth $d$ instead of $z$,
\begin{align}
	\px \int_h^{\zb} K \dd d &= \int_h^{\zb} \px K \dd d
\end{align}
The limits of the integral have been swapped because $h < \zb$ in depth coordinates. Now, because we express $K$ in terms of $d$ and $d = \zt - h$, where $\zt$ is the top of the aquifer,
\begin{align}
	\px K(d) &= \px K(\zt - h) \\[1ex]
	&= \frac{\p K}{\p d} \frac{\p (\zt - h)}{\p x} \\[1ex]
	&= \frac{\p K}{\p d} \left( \frac{\p \zt}{\p x} - \frac{\p h}{\p x} \right) \, .
\end{align}
Taking the integral of this expression,
\begin{align}
	\px \int_h^{\zb} K \dd d &= \int_h^{\zb} \frac{\p K}{\p d} \left( \frac{\p \zt}{\p x} - \frac{\p h}{\p x} \right) \dd d \\[1ex]
	&= \left( \frac{\p \zt}{\p x} - \frac{\p h}{\p x} \right) \int_h^{\zb} \frac{\p K}{\p d} \dd d \\[1ex]
	&= - \left( \frac{\p \zt}{\p x} - \frac{\p h}{\p x} \right) \int_{\zb}^h \frac{\p K}{\p d} \dd d \\[1ex]
	&= - \left( \frac{\p \zt}{\p x} - \frac{\p h}{\p x} \right) \int_{\zb}^h \frac{\p K}{\p h} \frac{\p h}{\p d} \dd d \\[1ex]
	&= - \left( \frac{\p \zt}{\p x} - \frac{\p h}{\p x} \right) \int_{\zb}^h \frac{\p K}{\p h} \dd h \, .
\end{align}
Finally, by the second fundamental theorem of calculus,
\begin{align}
	\px \int_{\zb}^h K \dd z &= \left( \frac{\p h}{\p x} - \frac{\p \zt}{\p x} \right) K(h) \label{dKintind}
\end{align}
This last expression is eminently useful because it allows the evaluation of Expression \ref{dKint} indirectly, in terms of the gradient of $\zt$ (time-independent), the gradient of $h$ (has to be evaluated anyway), and $K$ itself (should be easy and fast). This expression assumes that the bottom of the aquifer is at a uniform depth and that the hydraulic conductivity depends only on the depth. To confirm that Equation \ref{dKintind} holds, it was tested numerically for several different forms of $K(d)$, $\zt$, and $h$. One of these tests is shown in Figure \ref{fig:dKint}, with an explantion in the caption. With Equation \ref{dKintind}, the conservation equation becomes

\begin{equation}
	\boxed{ n(h) \frac{\p h}{\p t} = \frac{\p h}{\p x} \left( \frac{\p h}{\p x} - \frac{\p \zt}{\p x} \right) K(h) + \frac{\p^2h}{\p x^2} \int_{\zb}^h K \dd z + F } \, . \label{consnum}
\end{equation}

A solution algorithm will still have to evaluate the integral of $K$ numerically for each spatial node, but will not need to evaluate the gradient of its vertical integral directly. Evaluating these integrals could easily be the most computationally expensive part of a solver, so the indirect evaluation of $\nabla \int K$ could provide a huge efficiency boost.

All of that said, if the equation is being solved in a situation where the water table will intersect the surface or the bottom of the aquifer, discontinuities in the derivatives of $h$ are likely to cause problems for finite-difference and spectral methods. In this case, \textbf{a simple finite-volume method is probably the best choice}. With a finite-volume method, the flux in Equation \ref{Q} can be computed directly at cell edges and there is no need for Equation \ref{consnum} above. Assuming its poor stability properties can be overcome with an implicit time-stepper or by brute computational force, the simple midpoint finite-volume scheme works well because it requires the lowest order polynomial interpolation between cells but is still formally second order accurate. Accurate implicit methods with nonlinear systems can be tricky, so a fast, explicit solver may be necessary.

\section{Steady State}

First, to examine how the water table behaves under the Dupuit assumption and with significant vertical differences in the hydraulic conductivity, we look at steady state solutions. In steady state, the differential conservation law is
\begin{align}
	\nabla \cdot \left( \nabla h \int_{\zb}^h K \dd z  \right) &= -F \label{steady1int}
\end{align}
and in the simplest case where $\zb=0$ so that $T=h$,
\begin{align}
	\nabla \cdot \left( \Kbar h \nabla h \right) &= -F \, . \label{steady1}
\end{align}
Because $\nabla (h^2) = 2 h \nabla h$, this can immediately be rewritten as
\begin{align}
	\nabla \cdot \left[ \Kbar \nabla (h^2) \right] = -2F \, . \label{steady2}
\end{align}
The switch from $h$ to $h^2$ can be useful in the steady case because the equation can be manipulated or solved with respect to $h^2$ alone, forgetting about $h$. The transient equation (Equation \ref{cons2}) can't be solved in terms of $h^2$ because, even though $\p h/\p t$ could be expressed in terms of $h^2$, the transformation brings in other factors and doesn't simplify things on the whole. Perhaps using $h^2$ instead of $h$ in the spatial derivatives of the transient equation would simplify a numerical solver though.

\subsection{Constant Hydraulic Conductivity}

In the simplest case, $K$ is constant and
\begin{equation}
	\Kbar = \frac{K}{T} \int_{\zb}^h \dd z = \frac{K}{T} (h - \zb) \, .
\end{equation}
If the bottom of the aquifer is at a constant reference height of zero, then $\zb = 0$, $T=h$, $\Kbar = K$, and Equation \ref{steady2} becomes
\begin{equation}
	\nabla^2 (h^2) = -\frac{2F}{K} \, . \label{constK}
\end{equation}
This is a well-known result \cite{Bear72,Polub62} and is the simplest representation of a primarily horizontal groundwater system. If $F$ is known, the Laplacian can be solved numerically for $h^2$ with relative ease. It is, however, severely restricted by the assumption that $K$ is constant (both horizontally and vertically). It's not appropriate for deep or heterogeneous aquifers, where $K$ varies by significant factors. Nevertheless, it illustrates the fact that an unconfined aquifer with nearly horizontal flow is mediated by the balance of two parameters. These are the source term $F$, which is the net recharge, and the hydraulic conductivity $K$, which defines the system's drainage agility.

\subsection{Linear Hydraulic Conductivity}

The simplest vertically variable hydraulic conductivity is probably one that varies linearly with depth. If the ground surface is represented by some general horizontal function or data $s(x,y)$, the depth $d$ of the water table is
\begin{equation}
	d(x,y) = s(x,y) - h(x,y) \, . \label{depth}
\end{equation}
For an aquifer with uniform conductivity at the surface and zero conductivity at its maximum depth $D$, the linear conductivity function is
\begin{align}
	K(d) &= K_0 (1 - d/D) \\[1ex]
	K(h) &= K_0 [1 + (h - s)/D] \,.
\end{align}
From Equation \ref{steady1int}, we need the vertical integral of this conductivity for the governing equation.
\begin{align}
	\int_{\zb}^h K(h') &= K_0 \int_{\zb}^h \left( 1 + \frac{h' - s}{D} \right) \dd h' \\[1em]
	&= K_0 \int_{\zb}^h \left( 1 - \frac{s}{D} + \frac{h'}{D} \right) \dd h' \\[1ex]
	&= \frac{K_0}{D}\left[ (h - \zb)(D - s) + \frac{h^2 + \zb^2}{2} \right]
\end{align}
Plugging this in,
\begin{equation}
	\nabla \cdot \left( \left[ (h - \zb)(D - s) + \frac{h^2 + \zb^2}{2} \right] \nabla h \right) = -\frac{F D}{K_0} \, .
\end{equation}
This equation could be treated numerically if functions for $\zb$ and $s$ are known, but the simplest case that captures the linear variation of $K$ is where the aquifer thickness is uniform and $\zb=0$ everywhere. In one dimension, this means $\zb$ and $s$ are just horizontal lines and the aquifer is a rectangle. In this case, the equation above boils down to
\begin{equation}
	\nabla \cdot \left( h^2 \nabla h \right) = -\frac{2 F D}{K_0} \, .
\end{equation}
This looks quite similar to Equation \ref{constK}, where the conductivity was constant. Further, because $\nabla h^3 = 3 h^2 \nabla h$, we can rewrite the equation in terms of $h^3$,
\begin{equation}
	\boxed{ \nabla^2  (h^3) = -\frac{6 F D}{K_0} } \, , \label{linearK}
\end{equation}
which looks even more like Equation \ref{constK} and is easy to solve numerically (or analytically if $F$ and the boundary conditions are simple enough). The length dimension of the left side of the above equation has increased in order. Correspondingly, the length dimension of the right side has as well, because the aquifer depth $D$ now appears there.

\subsection{Monomial Hydraulic Conductivity}

In the linear conductivity case above, the steady conservation equation in a uniform thickness aquifer with linearly decreasing conductivity was shown to be solvable in terms of $\nabla^2 h^3$. A similar, simple result is available for conductivity functions decaying more quickly with depth, so long as the aquifer is still just a rectangle.

\begin{figure}[ht]
	\centering
	\includegraphics[scale=0.7]{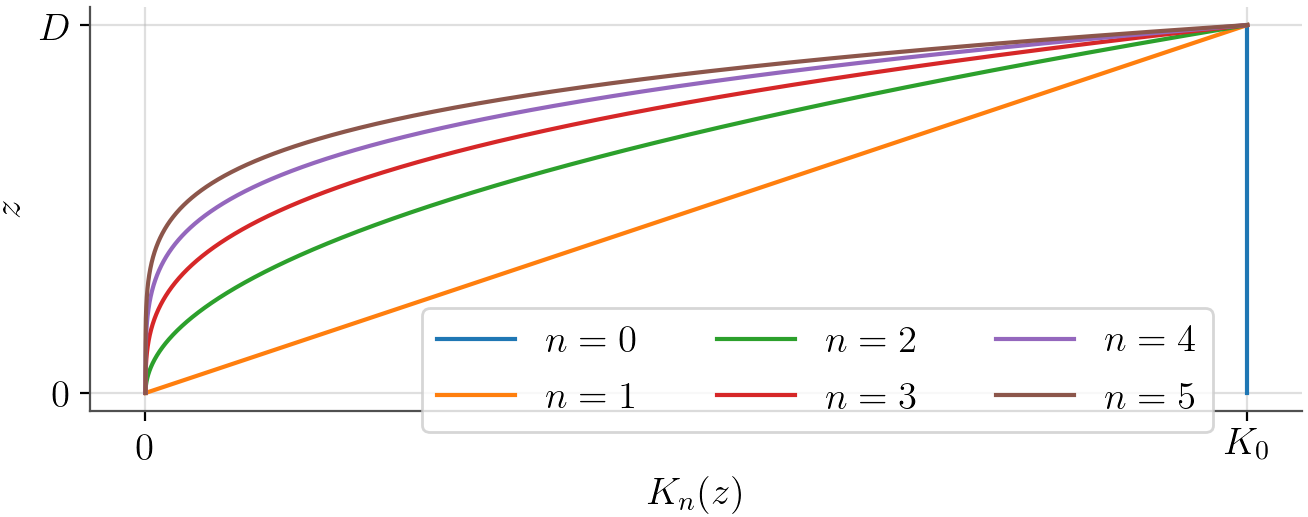}
	\caption{Monomial hydraulic conductivity profiles for orders 0 to 5. As the order increases, deeper parts of the aquifer become less conductive.}
	\label{fig:monomialK}
\end{figure}

If the n\ts{th} order conductivity function $K_n$ is
\begin{equation}
	K_n(z) = \frac{K_0}{D^n} z^n \, ,
\end{equation}
we have a family of simple functions where $K_n(0) = 0$ and $K_n(D) = K_0$ for all $n$. The zeroth order function is a constant, the first order function is a line, and higher order functions decay more with depth. Figure \ref{fig:monomialK} shows conductivity profiles for general $K_0$ and $D$, for $n$ up to 5. Higher order conductivity functions make deep parts of the aquifer less conductive.

The integral of each monomial conductivity function is easy to evaluate for substitution into Equation \ref{steady1int}. We've assumed the aquifer bottom is at $z = 0$, so the lower bound of the integral is zero.
\begin{align}
	\int_0^h K_n(z) \, \dd z &= \frac{K_0}{D^n} \int_0^h z^n \, \dd z \\[1ex]
	&= \frac{K_0}{D^n (n+1)} h^{n+1}
\end{align}
Plugging this into the governing equation (Equation \ref{steady1int}),
\begin{align}
	\nabla \cdot \left( \frac{K_0}{D^n (n+1)} h^{n+1} \nabla h \right) &= -F \\[1ex]
	\nabla \cdot \left( h^{n+1} \nabla h \right) &= -\frac{F D^n}{K_0}(n+1) \, .
\end{align}
Finally, we can apply the chain rule trick for general $n$,
\begin{equation}
	\nabla h^{n+2} = (n+2) h^{n+1} \nabla h \, ,
\end{equation}
and rearrange, yielding
\begin{align}
	\nabla \cdot \left( \frac{\nabla h^{n+2}}{n+2} \right) &= -\frac{F D^n}{K_0}(n+1)
\end{align}
\begin{equation}
	\boxed{ \nabla^2 (h^{n+2}) = -\frac{F D^n}{K_0}(n+1)(n+2) } \, . \label{monomial}
\end{equation}
The equation above shows that, for a general monomial conductivity function that is zero at the base of a uniform depth aquifer, the steady, Dupuit approximated water table is governed by the Laplacian of a power of $h$ two orders higher than that of the conductivity function. As a quick check, plugging in $n=0$ gives Equation \ref{constK} for constant $K$ and plugging in $n=1$ gives Equation \ref{linearK} for linear $K$.

\begin{figure}[ht]
	\centering
	\includegraphics[scale=0.8]{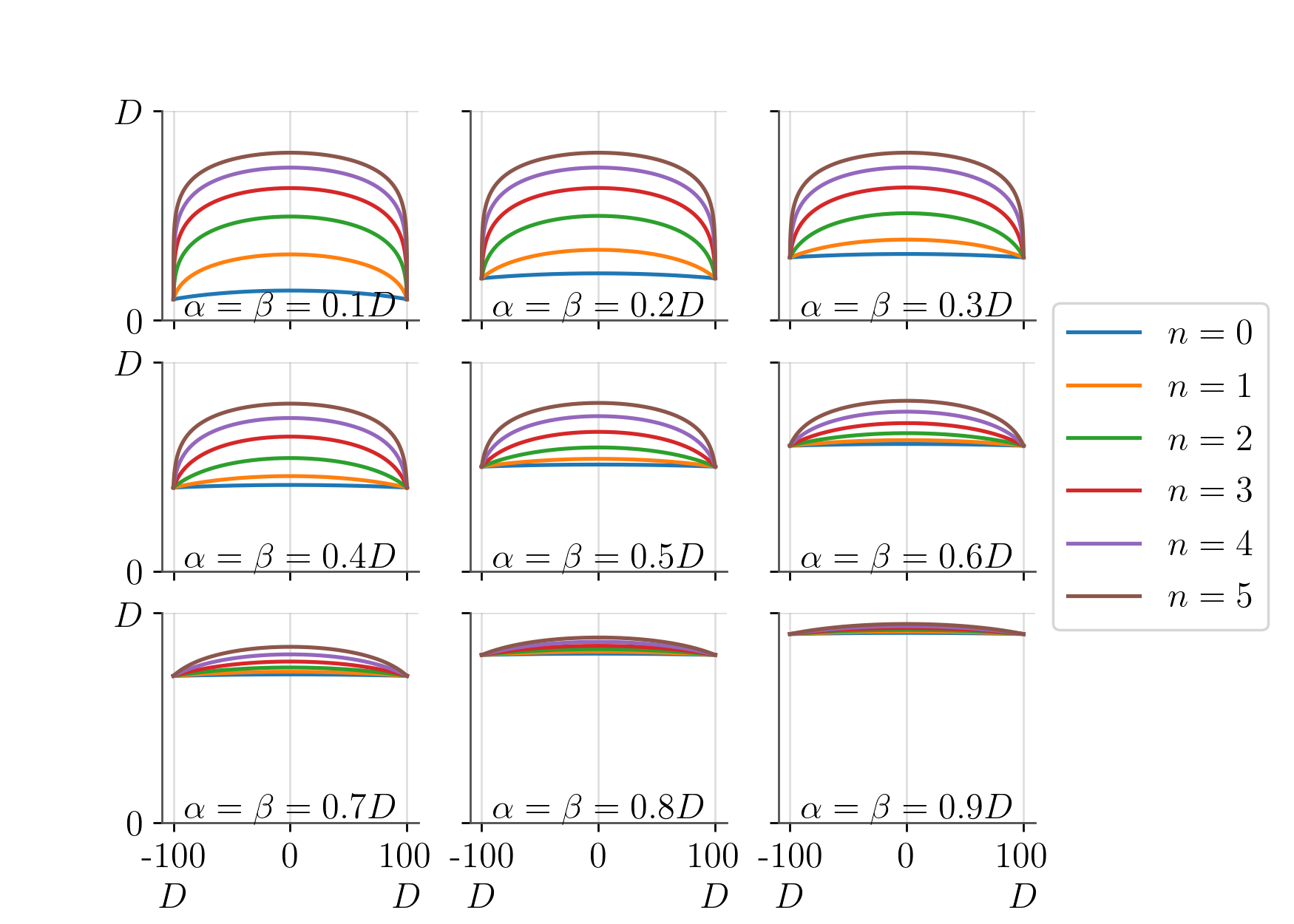}
	\caption{Analytical solutions to Equation \ref{monomial1d} for order $n$ up to five and for several different boundary water table values. Where the hydraulic conductivity is low, the hydraulic gradient is steep. For more rapidly decaying conductivity profiles, this pushes the water table up toward the surface. When the boundary values are closer to the surface, this effect is less pronounced.}
	\label{fig:monomialKsols}
\end{figure}

In one dimension, Equation \ref{monomial} is
\begin{equation}
	\frac{\p^2 (h^{n+2})}{\p x^2} = -\frac{F D^n}{K_0}(n+1)(n+2) \, . \label{monomial1d}
\end{equation}
If $F$ is a constant, then the whole right side of the equation above is constant. We can wrap it up into a single symbol,
\begin{equation}
	\Theta(n) \equiv \frac{F D^n}{K_0}(n+1)(n+2) \, ,
\end{equation}
and write the analytical solution of Equation \ref{monomial1d} over a horizontal length $L$ and boundary values of $\alpha$ and $\beta$,
\begin{align}
	h(-L/2) &= \alpha  &  h(L/2) &= \beta
\end{align}
\begin{align}
	c_1 &\equiv \frac{\beta^{n+2} - \alpha^{n+2}}{L} &  c_2 &\equiv \frac{1}{2}\left( \alpha^{n+2} + \beta^{n+2} + \frac{\Theta(n) L^2}{4} \right)
\end{align}
\begin{align}
	h(x) &= \left[ -\frac{\Theta(n)}{2} x^2 + c_1 x + c_2 \right]^{\frac{1}{n+2}} \, .
\end{align}
The constants $c_1$ and $c_2$, like $\Theta(n)$, are defined just for convenience. Figure \ref{fig:monomialKsols} shows the curves defined by the analytical solutions above for values of $n$ up to five and for nine different boundary values, each with $F = 1 \times 10^{-6}$. The curves are essentially shown in non-dimensional form, with all lengths expressed in terms of $D$.

\subsection{Polynomial Hydraulic Conductivity}

The same manipulations can be done with a sum of monomial conductivities (polynomial conductivity),
\begin{align}
	K(z) = K_0 \sum_{n=0}^N a_n \frac{z^n}{D^n} \, .
\end{align}
The only additional requirement is that $\sum a_n = 1$, ensuring that $K(D) = K_0$. In this case, the integral of $K$ over the saturated zone of the uniform depth aquifer is
\begin{align}
	\int_0^h K_0 \sum_{n=0}^N a_n \frac{z^n}{D^n} \, \dd z &= K_0 \sum_{n=0}^N \int_0^h a_n \frac{z^n}{D^n} \, \dd z \\[1ex]
	&= K_0 \sum_{n=0}^N a_n \frac{h^{n+1}}{D^n(n+1)} \, .
\end{align}
Putting the expression above into Equation \ref{steady1int}, shuffling things around, and applying the chain rule,
\begin{align}
	\nabla \cdot \left( \nabla h \sum_{n=0}^N a_n \frac{h^{n+1}}{D^n(n+1)} \right) &= -\frac{F}{K_0}
\end{align}
\begin{equation}
	\boxed{ \sum_{n=0}^N a_n \frac{\nabla^2 (h^{n+2})}{D^n (n+1) (n+2)} = -\frac{F}{K_0} } \, .
\end{equation}
I'm not sure how this kind of expression, with Laplacians of different powers of $h$, would be solved efficiently. One could just throw it into a big root finder if need be.

\subsection{Exponential Hydraulic Conductivity}

If the hydraulic conductivity decays exponentially over with depth,
\begin{align}
	K(d) &= K_0 e^{-d/H} \, ,
\end{align}
where $H$ is the decay constant. With variable topography $s$, the depth at $z$ is $d = s - z$.
\begin{align}
	K(z) &= K_0 e^{(z - s)/H}
\end{align}
For this case, the vertically integrated conductivity in the saturated zone is
\begin{align}
	K_0 \int_{\zb}^h e^{(z - s)/H} \dd z &= K_0 H \left[ e^{(h - s)/H} - e^{(\zb - s)/H} \right]
\end{align}
Plugging this into Equation \ref{steady1} and letting the factor of $h$ inside the divergence cancel with the one in $\Kbar$,
\begin{align}
	\nabla \cdot \left( H \left[ e^{(h - s)/H} - e^{(\zb - s)/H} \right] \nabla h \right) &= -\frac{F}{K_0} \label{expK1} \\[1ex]
	\nabla \cdot \left( \left[ e^{(h - s)/H} - e^{(\zb - s)/H} \right] \nabla h \right) &= -\frac{F}{H K_0} \, . \label{expK2}
\end{align}
The equation above is, of course, similar to Equation \ref{constK} in many ways. The source term on the right is controlled by the balance between $F$ and $K$, as before, but now also the decay constant (or scale height, if you like) $H$.

As a quick consistency check, we expect the equation above to look very close to Equation \ref{constK} when $H \gg D$ because the conductivity will be nearly constant over the depth of the aquifer. Without going through the algebra, WolframAlpha tells me that
\begin{equation}
	\lim_{H\to\infty} H \left[ e^{(h - s)/H} - e^{(\zb - s)/H} \right] = h - \zb
\end{equation}
If we put the bottom of the aquifer at zero ($\zb = 0$) as usual, then Equation \ref{expK1} becomes
\begin{equation}
	\nabla \cdot \left(h \nabla h \right) = -\frac{F}{K_0} \, ,
\end{equation}
which is identical to the constant $K$ case (Equation \ref{constK}) after applying the chain rule trick, $\nabla h^2 = 2h\nabla h$. So, we have consistency.

Looking at Equation \ref{expK2} again, we could find a way to get a numerical solution that captures variable $F$, $s$, and $\zb$ if we wanted to. Alternatively, we can make some simplifying assumptions or idealizations to better understand the behavior of the water table when conductivity decays exponentially with depth. With monomial conductivity profiles, the equations were simplified by putting the bottom of the aquifer at zero because the conductivity was designed to be zero there. Exponential conductivity doesn't afford that opportunity because it's never zero, but we can simplify by sending the aquifer bottom to negative infinity, $\zb \rightarrow \infty$. This removes one of the exponential terms in Equation \ref{expK2}. It might seem preposterous to invoke an infinitely deep aquifer, but its equivalent to asserting that the aquifer depth is more than a few times larger than $H$, in which case the second exponential in Equation \ref{expK2} is small enough to neglect. It's also sufficient to say that the water table never goes so deep that the bottomless aquifer becomes a problem.

After using infinity to get rid of one exponential term in Equation \ref{expK2}, we put the aquifer surface at zero ($s = 0$) to simplify the remaining exponential.
\begin{equation}
	\nabla \cdot \left( e^{h/H} \nabla h \right) = -\frac{F}{H K_0} \label{expK3}
\end{equation}
Now, more chain rule!
\begin{align}
	\nabla e^{h/H} &= \frac{1}{H} e^{h/H} \nabla h \\[1em]
	e^{h/H} \nabla h &= H \nabla e^{h/H}
\end{align}
Plugging the relation above into Equation \ref{expK3},
\begin{align}
	\nabla \cdot \left( H \nabla e^{h/H} \right) &= -\frac{F}{H K_0}
\end{align}
\begin{equation}
	\boxed{ \nabla^2 \left( e^{h/H} \right) = -\frac{F}{H^2 K_0} }
\end{equation}
The Laplacian is easily solved in terms of the exponential (numerically for general $F$ or analytically for simple $F$), and the water table is recovered with
\begin{equation}
	h = H \ln\left( e^{h/H} \right) \, .
\end{equation}
For the simple, one-dimensional case with constant $F$, the analytical solution with Dirichlet boundary conditions and $\Phi(H) \equiv F/H^2 K_0$ is
\begin{align}
	h(-L/2) &= \alpha  &  h(L/2) &= \beta
\end{align}
\begin{align}
	c_1 &\equiv \frac{e^{\beta/H} - e^{\alpha/H}}{L} &  c_2 &\equiv \frac{1}{2}\left( e^{\alpha/H} + e^{\beta/H} + \frac{\Phi(H) L^2}{4} \right)
\end{align}
\begin{align}
	h(x) &= H \ln \left[ -\frac{\Phi(H)}{2} x^2 + c_1 x + c_2 \right] \, ,
\end{align}
for a domain of width $L$, centered on zero, with values of $\alpha$ and $\beta$ on the boundaries.

\section{Transient}

Rewriting Equation \ref{cons3}, the general form of the transient conservation law,
\begin{equation}
	n(h) \frac{\p h}{\p t} = \nabla \cdot \left( \nabla h \int_{\zb}^h K \dd z \right) + F \, .
\end{equation}
Using a fancy $\K$ for the integral in that equation to distinguish it from the vertical average $\Kbar$ and pointy brackets $\langle\rangle$ for vectors,
\begin{align}
	n(h) \frac{\p h}{\p t} &= \nabla \cdot \left( \K \nabla h \right) + F \\[1ex]
	&= \left\langle \px , \py \right\rangle \cdot \K \left\langle \frac{\p h}{\p x} , \frac{\p h}{\p y} \right\rangle + F \\[1ex]
	&= \px \K \frac{\p h}{\p x} + \py \K \frac{\p h}{\p y} + F \\[1ex]
	&= \K \frac{\p^2 h}{\p x^2} + \frac{\p h}{\p x} \frac{\p \K}{\p x} + \K \frac{\p^2 h}{\p y^2} + \frac{\p h}{\p y} \frac{\p \K}{\p y} + F \\[1ex]
	&= \K \nabla^2 h + ( \nabla \K \cdot \nabla h) + F \, .
\end{align}
This is a general form of Equation \ref{consnum}, which we simplified by assuming that $K$ depends only on depth and that the thickness of the aquifer is constant. Generalizing Equation \ref{consnum} to two dimensions then,
\begin{equation}
	n(h) \frac{\p h}{\p t} = K(h) [\nabla \cdot (\nabla h - \nabla \zt)] + \K \nabla^2 h + F \, . 
\end{equation}

\end{document}